# Analyzing Highly Volatile Driving Trips Taken by Alternative Fuel Vehicles

Mohsen Kamrani
Department of Civil & Environmental Engineering
The University of Tennessee
mkamrani@vols.utk.edu

Ramin Arvin
Department of Civil & Environmental Engineering
The University of Tennessee
rarvin@vols.utk.edu

Asad J. Khattak, Ph.D.
Department of Civil & Environmental Engineering
The University of Tennessee
akhattak@utk.edu

## TRB PAPER # 18-00980

**Analyzing Highly Volatile Driving Trips Taken by Alternative Fuel Vehicles**

Mohsen Kamrani, Ramin Arvin, Asad J. Khattak
The University of Tennessee, Knoxville

**Abstract –** Volatile driving, characterized by fluctuations in speed and accelerations and aggressive lane changing/merging, is known to contribute to transportation crashes. To fully understand driving volatility with the intention of reducing it, the objective of this study is to identify its key correlates, while focusing on highly volatile trips. First, a measure of driving volatility based on vehicle speed is applied to trip data collected in the California Household Travel Survey during 2012-2013. Specifically, the trips containing driving cycles (N=62839 trips) were analyzed to obtain driving volatility. Second, correlations of volatility with the trip, vehicle, and person level variables were quantified using Ordinary Least Squares and quantile regression models. The results of the 90th percentile regression (which distinguishes the 10% highly volatile trips from the rest) show that trips taken by pickup trucks, hatchbacks, convertibles, and minivans are less volatile when compared to the trips taken by sedans. Moreover, longer trips have less driving volatility. In addition, younger drivers are more volatile drivers than old ones. Overall, the results of this study are reasonable and beneficial in identifying correlates of driving volatility, especially in terms of understanding factors that differentiate highly volatile trips from other trips. Reductions in driving volatility have positive implications for transportation safety. From a methodological standpoint, this study is an example of how to extract useful (volatility) information from raw vehicle speed data and use it to calm down drivers and ultimately improve transportation safety.

Keywords:  Driving volatility, Aggressive Driving, Quantile Regression, Travel Survey

## INTRODUCTION

The availability of precise data on driving speed, acceleration, etc. has led to the creation of different methods to process and analyze driving styles (*1; 2*). Specifically, aggressive driving is of importance due to its correlation with risk of crashes. Although some thresholds can be defined to specify the moments of aggressive driving, the new concept of "driving volatility" that captures the level of variations in several driving elements such as speed, acceleration, and jerk, provides a more objective way of quantifying aggressive driving. It is noteworthy that driving volatility was found to be positively associated with crash frequency (*3*) and therefore, one of the ways to reduce the risk of crashes is to reduce driving volatility i.e. to drive smoother and calmer. In this regards, understanding the correlates of driving volatility would be beneficial. California national household survey driving cycle data provides an opportunity for creating trip-based driving volatilities database. The data collected in the California Household Travel Survey (CHTS) (*4*) that conducted travel household survey for California Department of Transportation is the largest single regional or statewide household travel survey ever conducted in the U.S. The survey was conducted during January 2012 through January 2013. The driver-cycles file that contains second-by-second data for each trip recorded by in-vehicle GPS (Global Positioning System) and OBD (On-Board Diagnostics) devices (54 million driving records) was used for calculation of driving volatility. Driving volatility data are integrated with the vehicle, trip, and demographic variables to be used in statistical models where driving volatility is treated as a response variable rather an independent variable to fulfill the main objective of this paper: To investigate the correlates of highly volatile trips as a first step of taking countermeasures to reduce driving aggressiveness.

By introducing a new way of capturing variation in the driving speed, this study contributes to how meaningful and useful information can be extracted and be used as the measure of driving volatility.

## LITERATURE REVIEW

In the current literature, the term "aggressive" driving is being used for describing the behaviors that threaten the driver and other vehicles' safety. In the United States, 56 percent of the fatality crashes from 2003 through 2007 are due to the aggressive driving behaviors such as speeding, reckless driving, failure to obey the stop sign and control devices, failure to yield the right of way, and improper turning (*5*). Various definitions are presented for defining the drivers' aggressiveness. Lajunen et al. (*6*) defined the aggressive driving as "any form of driving behavior that is intended to injure or harm other road users physically or psychologically." A wide range of behaviors can be categorized as the aggressive behavior from flashing light, verbal threats, tailgating, to the most extreme aggressiveness such as physical attacks (*7*). Some studies (*8; 9*) attempt to model the influence of aggressive driving on the injury severity. Tasca (*10*) is one of the first studies characterizing aggressive driving, defining aggressive driving as "it is deliberate, likely to increase the risk of collision and is motivated by impatience, annoyance, hostility and/or attempt to save time."

On the other hand, focusing on the instantaneous driving behavior, aggressive driving can be described using vehicle kinematics such as vehicle's speed, acceleration, and jerk. There are wide definitions for instantaneous aggressive driving in the literature. Most of the studies used common kinetic characteristics of the vehicle such as speed, longitudinal/lateral acceleration, and jerk for quantifying the aggressiveness or deviation from normal behavior (*11; 12*). Maximum acceleration/deceleration is one of the most favorable variables for describing aggressiveness. In the urban environment, Kim et al. (*13*) suggested the threshold for aggressive and extreme aggressive driving (1.47 m/s$^2$ and 2.28 m/s$^2$ respectively). For the city journeys, De Vlieger (*14*) defined various thresholds for different driving style among which aggressive driving is defined as a range of 0.85 to 1.10 m/s$^2$. Han et al. (*15; 16*) defined different thresholds varying with the vehicle speed to take into account the variation in the driving behavior under various driving conditions. Jerk, the acceleration rate change with respect to the time, is one of the measures being used for evaluation of the drivers' aggressiveness. Murphy et al. (*17*) used vehicular jerk to classify styles of aggressive driving using the ratio of the standard deviation to the mean for the jerk. Baghdadi et al. (*18; 19*) used jerk to determine the accident-prone drivers. Also, aggressive drivers associated with significantly higher values of the jerk (*20*).

Recently, different studies tried to capture driving behavior using "driving volatility" (*3; 21*). Volatility measures that capture the extent of variations in driving, especially hard accelerations/braking, jerky maneuvers, and frequent switching between different driving regimes (*22*) can be used to quantify the driving style. In this regard, driving cycle data from California national household survey could be for creating trip-based driving volatilities for drivers.

## METHODOLOGY

### Driving Volatility

As discussed earlier in the introduction, volatility measures can be used to quantify driving variation. Different measures are applied by researchers to quantify the driver volatility (*3; 23*). In this paper, Time Varying Stochastic Volatility is used which is a new measure of driving volatility. This measure provides a tool to time-varying stochastic volatility measure (*24*). This measure is able to capture the instantaneous changes in the driving speed of the vehicle. Large values of the volatility imply the abnormal variability in the driving behavior of the selected trip. The volatility can be calculated as (*21; 24*):

$$Driving\ Volatility = \sqrt{\frac{1}{n-1}\sum_{i=1}^{n}(r_i - \bar{r})^2} \quad \text{from } i = 1 \text{ to } n \tag{1}$$

Where

$$r_i = \ln\left(\frac{x_i}{x_{i-1}}\right) * 100 \tag{2}$$

where $x_i$ and $x_{i-1}$ are the current and previous speed respectively, *ln* is the natural logarithm, and $n$ is the number of observations (trip length).

**Modeling Approach**

As instantaneous driving decisions (and the volatility therein) are highly context (driver) specific, there is a real possibility that the correlations may be heterogeneous in nature, due to several observed and unobserved factors (*25; 26*). In this section, Ordinary Least Squares (OLS) regression and quantile regression will be discussed as a basis for the analysis. Especially the quantile regression is suitable for the studying highly volatile trips (90[th] percentile). In the quantile regression method, using different quantile points, it takes full advantages of the dataset to perform regression. Therefore, in the quantile regression, the behavior of each quantile is described by the specific function. There are several benefits of the quantile regression over OLS estimates.

1- In the quantile regression, there is no assumption of the existence of moment function (*27; 28*).
2- The results of quantile regression are robust, and the model will not be affected by the outlier, skewness, and heterogeneity in the dependent variable (*29*).
3- While in the OLS, the random error term is subjected to independent and identically distributed (IID) and normal distribution, in the quantile regression, there is no need for distributional assumption (*30*).

*OLS Model*

The OLS model is:

$$y_i = \beta_0 + \sum_{j=1}^{n}\beta_j x_{ij} + \varepsilon_i \tag{3}$$

where
$y_i$: dependent variable, here is volatility of driver *i, i=1,2,…, m;*
$\beta_0$: intercept of the model;
$\beta_j$: coefficient of independent variable of *j=1,2,…,n;*
$x_{ij}$: value of independent variable *j* for the driver *i;*
$\varepsilon_i$: estimation error for the volatility of driver *i;*

In OLS, it is assumed that the error term ($\varepsilon_i$) is normally distributed with the mean of zero and a finite variance. To estimate the coefficients of the model, the mean squared error, which is shown in the following equation, should be minimized:

$$\sum_{i=1}^{m}\left(y_i - \beta_0 - \sum_{j=1}^{n}\beta_j x_{ij}\right)^2 \tag{4}$$

The estimated intercept, $\beta_0$, and coefficients of independent variables, $\beta_j$, are denoted by $\widehat{\beta_0}$ and $\widehat{\beta_j}$, respectively. The results of the OLS model intuitively interpret the relationship between independent variables and dependent variables; one increase in an independent variable of $j$ is leading to $\beta_j$ change in the mean of the drivers' volatility.

*Quantile Regression*

OLS approach is summarizing the average relationship between independent variables and the dependent variable which might not be interested when we are intended to describe the relationship at different points in the distribution of the dependent variable. Focusing on the mean effects might lead to inaccurate estimation or omission of significant independent variables (*31*). To obtain a better picture of the distribution of drivers' volatility, quantile regression is more appropriate by modeling the relationship of any quantile with independent variables. This method is firstly introduced by Koekner and Bassett (*32*).

In this method, we assume that the value of $\varepsilon_i$ conditional on the independent variables in the $q^{th}$ quantile is zero. While in the OLS approach the mean squared error is minimized, in the quantile regression, the objective function is minimizing the sum that gives asymmetric penalties $(1-q)|\varepsilon_i|$ for over-prediction and $q|\varepsilon_i|$ for under-prediction; where q is denoting the quantile point of the outcome. The prediction error is given by:

$$\varepsilon_i^q = \hat{\beta}_0^q - \sum_{j=1}^{n}\hat{\beta}_j^q x_{ij} \tag{5}$$

where $\hat{\beta}_0^q$ is the estimated intercept of the model for quantile $q$ (0<$q$<1); $\hat{\beta}_j^q$ is the estimated coefficient of independent j at quantile point $q$. By considering trip volatility as dependent variable, all observations are divided into two groups based on the volatility related to the quantile value. In the first category, all observations have higher volatility than the specified quantile. The other group has lower volatility values than the specified quantile. Within each group, the total absolute deviation between driver volatility and specific quantile value can be calculated. The objective function is defined by weighting the average of total deviation within each category. To estimate $\hat{\beta}_0^q$ and $\hat{\beta}_j^q$, the following objective function should be minimized:

$$\sum_{i:y_i\geq\beta_0^q+\sum_{j=1}^{n}\beta_j^q x_{ij}}^{n} q\left|y_i - \beta_0^q + \sum_{j=1}^{n}\beta_j^q x_{ij}\right| + \sum_{i:y_i<\beta_0^q+\sum_{j=1}^{n}\beta_j^q x_{ij}}^{n}(1-q)\left|y_i - \beta_0^q + \sum_{j=1}^{n}\beta_j^q x_{ij}\right| \tag{6}$$

where $y_i$ is the volatility of driver $i$, $i$=1, 2,…,$n$; and $x_{ij}$ is the value of independent variable $j$ for driver $i$. Different statistical software packages are providing quantile regression models. In this study, we used QREG package in STATA.

## DATA

In this study, we chose driving behavioral data collected in the California Household Travel Survey (CHTS) (*4*) that conducted travel household survey for California Department of Transportation. The survey is the largest single regional or statewide household travel survey ever conducted in the U.S. The survey was conducted during January 2012 through January 2013.

This study used Person, Vehicle, Trip, and driver-cycles data. The driver-cycles file contains second-by-second data for each trip recorded by in-vehicle GPS (Global Positioning System) and OBD (On-Board Diagnostics) devices (54 million driving records). The dataset contains the travel date and time, speed, acceleration, latitude, and longitude information (geocode records were removed from the public release dataset). Using driving cycle data, historical finance volatility was calculated for each trip. Then, each trip is linked to the "Trip file" that contains a summary of all trips have taken by vehicles. This dataset provides information on vehicle and road such as distance traveled, average speed, acceleration, the number of stops, road grade, and elevation, etc. The final dataset is built by linking the trip information to the vehicle and household dataset using a household specific ID. The final dataset contains 62839 trips taken by 2677 vehicles. In the dataset, Hybrid, Plug-in hybrid electric, CNG, and electric vehicle are coded as Alternative Fuel Vehicles (AFV).

## RESULTS

### Descriptive Statistics

In Table 1, descriptive statistics of key variables are shown. Based on the table, values seem reasonable, and dataset was checked for errors. Missing values were replaced by the mean and only those variables used in the study are shown for the sake of brevity. Finance Historical Volatility is measured for 62839 trips. The average volatility of the trip is 13.081% (min=1.69, and max=39.694). Average speed was 27.608 mph ranging from 1.995 to 79.172 mph. With regard to vehicle characteristics, 22 percent of vehicles are AFV (hybrid, plug-in, hybrid electric, CNG, and electric vehicles), 86.6 percent have automatic transmissions, and the average number of engine cylinders is 6.71. Among 2677 drivers, the predominant group age is 50-59 years (31.6 percent), 55 percent of respondents are female, and 73 percent are employed, and 95 percent of vehicles are owned by the household.

Figure 1 illustrates the histogram of drivers' volatility of valid observations, N=62839. The histogram is showing that the distribution of the dependent variable is similar to the normal distribution and the OLS and quantile regression is appropriate for modeling the volatility.

**TABLE 1- Descriptive Statistics of the Data**

| Variable | N | Mean | Std. Dev. | Min | Max |
|---|---|---|---|---|---|
| Driving Volatility | 62839 | 13.081 | 3.644 | 1.69 | 39.694 |
| Distance traveled (miles) | 62839 | 8.295 | 14.902 | 0.002 | 342.484 |
| Travel time (min) | 62839 | 15.621 | 18.045 | 0.1 | 367.433 |
| Average speed (mph) | 62839 | 27.608 | 12.350 | 1.995 | 79.172 |
| Number of stops | 62839 | 3.601 | 3.16 | 1 | 116 |
| Std. dev. of road grade | 62839 | 1.517 | 1.154 | 0 | 7.931 |
| AFV (No=0, Yes=1) | 2677 | 0.22 | 0.417 | 0 | 1 |
| Body type | | | | | |
| Sedan | 2677 | 0.435 | 0.496 | 0 | 1 |
| SUV | 2677 | 0.194 | 0.396 | 0 | 1 |
| Pickup | 2677 | 0.122 | 0.327 | 0 | 1 |
| Coupe | 2677 | 0.057 | 0.231 | 0 | 1 |
| Convertible | 2677 | 0.014 | 0.118 | 0 | 1 |
| Hatchback | 2677 | 0.08 | 0.271 | 0 | 1 |
| Wagon | 2677 | 0.039 | 0.193 | 0 | 1 |
| Minivan | 2677 | 0.047 | 0.212 | 0 | 1 |
| Van | 2677 | 0.012 | 0.107 | 0 | 1 |
| Other | 2677 | 0.001 | 0.027 | 0 | 1 |
| Transmission | | | | | |
| Automatic | 2677 | 0.866 | 0.341 | 0 | 1 |
| Manual | 2677 | 0.1 | 0.3 | 0 | 1 |
| Both | 2677 | 0.035 | 0.183 | 0 | 1 |
| Vehicle Age (years) | 2677 | 8.03 | 4.725 | 0 | 52 |
| Number of cylinders | 2677 | 6.714 | 3.501 | 2 | 12 |
| Power train | | | | | |
| Front-wheel | 2677 | 0.538 | 0.499 | 0 | 1 |
| Rear-wheel | 2677 | 0.275 | 0.447 | 0 | 1 |
| Four-wheel | 2677 | 0.186 | 0.39 | 0 | 1 |
| Gender (male=0, female=1) | 2677 | 0.55 | 0.498 | 0 | 1 |
| Age (years) | | | | | |
| 10-19 | 2677 | 0.023 | 0.149 | 0 | 1 |
| 20-29 | 2677 | 0.071 | 0.256 | 0 | 1 |
| 30-39 | 2677 | 0.144 | 0.351 | 0 | 1 |
| 40-49 | 2677 | 0.217 | 0.412 | 0 | 1 |
| 50-59 | 2677 | 0.316 | 0.465 | 0 | 1 |
| 60-69 | 2677 | 0.192 | 0.394 | 0 | 1 |
| 70-79 | 2677 | 0.034 | 0.181 | 0 | 1 |
| 80-89 | 2677 | 0.004 | 0.061 | 0 | 1 |
| Ownership (owned=0,non-owned=1) | 2677 | 0.05 | 0.219 | 0 | 1 |
| Employment (Yes=0, No=1) | 2677 | 0.27 | 0.444 | 0 | 1 |

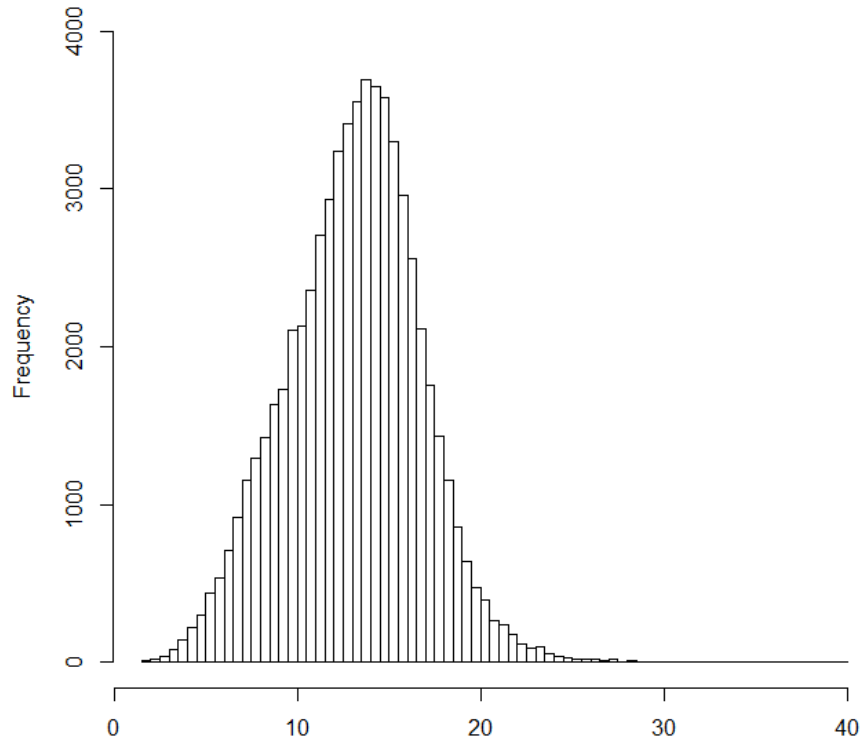

**FIGURE 1 Histogram of driving volatility**

## Modeling Results

The study is focused on the 90th percentile to address the contributing factors to highly volatile trips. Table 2 presents the OLS and quantile regressions results where trips volatility (%) is the dependent variable. Previous studies used volatility as one of the independent variables and suggested that the driving volatility is positively associated with higher risk of crashes. Consequently, the reduction of volatility would be helpful to reduce the crash risk and severity. Therefore, understanding the correlates of driving volatility might shed some light on what measures should be taken to reduce volatility or what groups are more volatile in terms of driving. While the OLS shows the direction and magnitude of the relationship between the driving volatility and independent variables, the quantile regressions provide the trend and change in the estimates at different percentiles of the dependent variable.

Looking at the variable "AFV" shows that for highly volatile trips (90th percentile), the trips taken by AFV are less volatile in comparison with trips taken by conventional vehicles controlling for other variables. The overall trend in Figure 2 suggests the estimates for AFV variable across the quantiles is different. From the figure, it can be concluded that for drivers who are already highly volatile, the effect of driving AFV in terms of calming their driving, would be greater as compared to a driver with low volatility. The reason that AFVs are less volatile might be attributed to their different engine, limited acceleration capability as compared to conventional vehicles or the existence of the fuel consumption optimization system.

It is intuitive to consider the body type as one of the contributing factors to the driving volatility. Vehicle body type affects the aerodynamic drag and maneuver capability of the vehicles. Under body variable in Table 2, the estimates of various body types as compared the sedan vehicles are provided. The results are very interesting in the sense that some estimates for highly volatile trips are in the opposite direction of the low volatile trips. The SUV body type estimates at 90th percentile show that highly volatile trips taken by SUVs are less volatile as compared to the sedans (although the estimate is statistically insignificant at 5% CI, possibly because of small sample size of highly volatile trips taken with SUV). However, for less volatile trips, the effect of driving an SUV is positive (compared with sedan). Figure 3 shows the concave trend of SUV estimates across the quantile. The estimate changes from positive to negative at around 80$^{th}$ percentile. Highly volatile trips taken by pickup trucks, hatchbacks, convertibles, and minivans are less volatile than trips taken by sedans. Figure 3 also shows how the effect of driving a coupe vehicle is larger in highly volatile trips as compared to low volatile ones. Overall, the results from the vehicle body types are reasonable and intuitive. Also, they are good examples of how quantile regressions can unveil the correlations not captured by OLS regression.

Other vehicle-related variables include transmission type, vehicle age, the number of cylinders and the type of the powertrain. It is interesting that manual vehicles are associated with less driving volatility when compared with automatic vehicles. However, for highly volatile trips, the estimate is not statistically significant. Older vehicles are less volatile than the new ones (again not significant for highly volatile trips). While the estimate for the number of cylinders variable is not statistically significant from the OLS regression, the quantile regression results (Table 2 and Figure 3) suggest that for up to 45$^{th}$ percentile of the driving volatility, the association is positive and it becomes negative afterward some of which are statistically insignificant. Moreover, the results indicate the volatility of rear-wheel vehicles as compared to the base (front-wheel vehicles) is not statistically significant. That means their volatility is not different across quantiles. However, the effect of all-wheel drive vehicles when compared the front-wheel drive vehicles is negative. It means trips taken by all-wheel drive vehicles are less volatile compared to trips taken by front wheel drive vehicles. Focusing on high volatile trips at 90$^{th}$ percentile, the effect considerably is less when compared to low volatile trips. In other words, the type of powertrain for already high volatile drivers is less compared to drivers with a low level of driving volatility.

Three trip-related variables are also included in the models: trip distance, the number of the stops during the trip and the standard deviation of the road grade. For highly volatile trips, as the trip distance increases for one mile, the driving volatility decreases by 0.12% while for low volatile trips the amount of decrease is 0.14% (for one mile increase in travel distance). Figure 2 shows the convex shape of the effect of travel distance on the volatility across the quantiles. The figure shows the effect of travel distance on low volatile drivers is greater as compared with highly volatile drivers. The overall results of trip distance are intuitive in the sense that shorter trips commonly have higher levels of driving volatility because they are involved with traffic, signal lights and thus more acceleration and deceleration events. In addition, the number of stops during a trip is positively correlated with driving volatility. For highly volatile trips, one additional stop during the trip is associated with 0.13% increase in the driving volatility while the effect for low volatile trips is 3.4 times larger. It should be noted that the number of the stops and trip distance could be highly correlated. However, this is not the case for this data, given the fact that their variance of inflation factor (VIF) are 1.09 and 1.06 respectively (0.27 Pearson correlation). The estimates for the standard deviation of the road grade during the trips suggests that the one unit increase in the road grade standard deviation (road grade unit is also in %) for high volatile trips is associated with 0.7% decline in the driving volatility. If we assume that higher standard deviation of the road grade means more upgrades and downgrades in the trips, the results are intuitive because in those situations drivers are more careful with controlling the vehicle and the variation of speed i.e. keeping their driving volatility as low as possible. The interesting part from these results is that the effect is greater for highly volatile trips (drivers) as compared with low volatile trips (see the concave shape of the estimate in Figure 2).

The final set of independent variables included in the models are demographics that are mainly used for controlling purposes. Still, they can provide some useful insight on how much the driving volatility differs among different groups. Looking at the categorical variable of age, the estimates show as the group age increases, the magnitude of reduction in driving volatility, compared to the base group, declines. Ownership variable results indicate that drivers drive more volatile if they do not own the vehicles. The magnitude of effect for highly volatile trips is greater than low volatile trips. It is interesting to note that when it comes to the employment status, the results indicate the non-employed drivers are more volatile than the employed ones. The employment estimate for highly volatile trips is not statistically significant. The graphical quantile regression results for discussed variables are presented in Figures 2 and 3.

**TABLE 2 OLS and Quantile Regression Results (N=62839)**

| Variable | OLS (mean) | 10th Percentile | 25th Percentile | 50th Percentile | 75th Percentile | 90th Percentile |
|---|---|---|---|---|---|---|
| | Coef. (t-value) | Coef. (t-value) | Coef. (t-value) | Coef. (t-value) | Coef. (t-value) | Coef. (t-value) |
| AFV | -0.328 (-10.9) | -0.221 (-6.4) | -0.305 (-10.5) | -0.354 (-11.4) | -0.360 (-8.0) | -0.391 (-5.8) |
| Body type (base: Sedan) | | | | | | |
| SUV | 0.031 (0.93) | 0.015 (0.4) | 0.068 (2.09) | 0.082 (2.36) | 0.056 (1.1) | -0.114 (-1.5) |
| Pickup | -0.668 (-15.51) | -0.734 (-14.8) | -0.592 (-14.28) | -0.612 (-13.89) | -0.662 (-10.3) | -0.779 (-8.0) |
| Coupe | 0.225 (4.15) | 0.056 (0.9) | 0.131 (2.51) | 0.161 (2.9) | 0.331 (4.1) | 0.308 (2.5) |
| Convertible | -0.074 (-0.75) | 0.306 (2.6) | 0.113 (1.18) | 0.155 (1.52) | -0.205 (-1.3) | -0.470 (-2.1) |
| Hatchback | -0.184 (-4.15) | -0.078 (-1.5) | -0.132 (-3.1) | -0.159 (-3.5) | -0.201 (-3.0) | -0.206 (-2.1) |
| Wagon | 0.319 (5.33) | 0.217 (3.1) | 0.242 (4.21) | 0.332 (5.42) | 0.285 (3.2) | -0.042 (-0.3) |
| Minivan | -0.171 (-3.3) | -0.293 (-4.9) | -0.290 (-5.8) | -0.194 (-3.65) | -0.189 (-2.4) | -0.257 (-2.2) |
| Van | -0.252 (-2.37) | -0.227 (-1.8) | -0.318 (-3.1) | -0.465 (-4.26) | -0.459 (-2.9) | -0.092 (-0.4) |
| Other | -0.399 (-0.91) | -1.413 (-2.8) | -0.745 (-1.76) | -0.289 (-0.64) | -0.145 (-0.2) | 0.476 (0.4) |
| Transmission (base: Automatic) | | | | | | |
| Manual | -0.304 (-7.45) | -0.315 (-6.7) | -0.354 (-9.0) | -0.321 (-7.6) | -0.363 (-5.9) | -0.139 (-1.5) |
| Both | 0.097 (1.55) | 0.112 (1.5) | 0.028 (0.4) | -0.029 (-0.4) | -0.090 (-0.9) | -0.002 (0.0) |
| Vehicle age | -0.008 (-3) | -0.012 (-3.8) | -0.011 (-4.1) | -0.005 (-1.9) | -0.008 (-1.9) | -0.009 (-1.6) |
| No. of cylinders | 0.001 (0.06) | 0.030 (2.1) | 0.020 (1.7) | -0.016 (-1.2) | -0.036 (-1.9) | -0.035 (-1.2) |
| Power train (base: Front-wheel drive) | | | | | | |
| Rear-wheel | -0.004 (-0.15) | 0.001 (0.0) | -0.025 (-0.9) | -0.002 (-0.1) | 0.045 (1.0) | 0.091 (1.4) |
| Four-wheel | -0.424 (-11.88) | -0.566 (-13.8) | -0.577 (-16.8) | -0.471 (-12.9) | -0.392 (-7.4) | -0.146 (-1.8) |
| Distance traveled (miles) | -0.139 (-177.2) | -0.226 (-250.8) | -0.235 (-310.2) | -0.219 (-271.7) | -0.176 (-150.9) | -0.123 (-69.9) |
| No. of stops | 0.280 (76.52) | 0.453 (107.8) | 0.448 (126.9) | 0.372 (99.1) | 0.244 (44.9) | 0.134 (16.2) |
| Std. dev. of road grade | -0.704 (-71.32) | -0.554 (-49.0) | -0.493 (-51.9) | -0.506 (-50.1) | -0.601 (-40.9) | -0.719 (-32.4) |
| Gender (Female=1) | 0.070 (2.92) | 0.016 (0.6) | 0.048 (2.1) | 0.071 (2.8) | 0.078 (2.1) | -0.003 (0.0) |
| Age (base: 10-19 years)  the actual range is 16-19 years | | | | | | |
| 20-29 | -0.132 (-1.63) | 0.063 (0.6) | -0.153 (-1.9) | -0.196 (-2.3) | -0.149 (-1.2) | -0.139 (-0.7) |
| 30-39 | -0.225 (-2.97) | -0.140 (-1.6) | -0.199 (-2.7) | -0.232 (-3) | -0.243 (-2.1) | -0.249 (-1.4) |
| 40-49 | -0.214 (-2.88) | -0.120 (-1.4) | -0.217 (-3.0) | -0.257 (-3.3) | -0.290 (-2.6) | -0.344 (-2.0) |
| 50-59 | -0.392 (-5.33) | -0.255 (-3.0) | -0.411 (-5.8) | -0.429 (-5.7) | -0.444 (-4.0) | -0.481 (-2.9) |
| 60-69 | -0.586 (-7.87) | -0.412 (-4.8) | -0.560 (-7.8) | -0.623 (-8.1) | -0.688 (-6.2) | -0.810 (-4.8) |
| 70-79 | -0.599 (-6.28) | -0.586 (-5.4) | -0.748 (-8.1) | -0.692 (-7.1) | -0.618 (-4.3) | -0.570 (-2.6) |
| 80-89 | -0.987 (-4.03) | -0.508 (-1.8) | -0.774 (-3.3) | -0.840 (-3.3) | -1.688 (-4.6) | -2.139 (-3.8) |
| Ownership (non-owned=1) | 0.342 (7.08) | 0.199 (3.6) | 0.195 (4.2) | 0.252 (5.1) | 0.348 (4.8) | 0.457 (4.21) |
| Employment (no=1) | 0.154 (5.59) | 0.069 (2.2) | 0.070 (2.6) | 0.069 (2.4) | 0.073 (1.8) | 0.075 (1.21) |
| Constant | 14.844 (170.1) | 11.644 (116.4) | 12.965 (154.5) | 14.623 (163.9) | 16.733 (129.1) | 19.001 (97.0) |
| $R^2$ | 0.419 | 0.392[a] | 0.373[a] | 0.300[a] | 0.207[a] | 0.139[a] |
| Raw sum of deviations | NA | 40969.8 | 74478.1 | 90392.44 | 70381.46 | 39473.44 |
| Min. sum of deviations | NA | 24905.5 | 46709.85 | 63256.52 | 55845.05 | 33972.14 |

[a] Represents pseudo-$R^2$ for quantile regression is calculated as *pseudo-$R^2$=1-(sum of deviations/raw sum of deviations)*

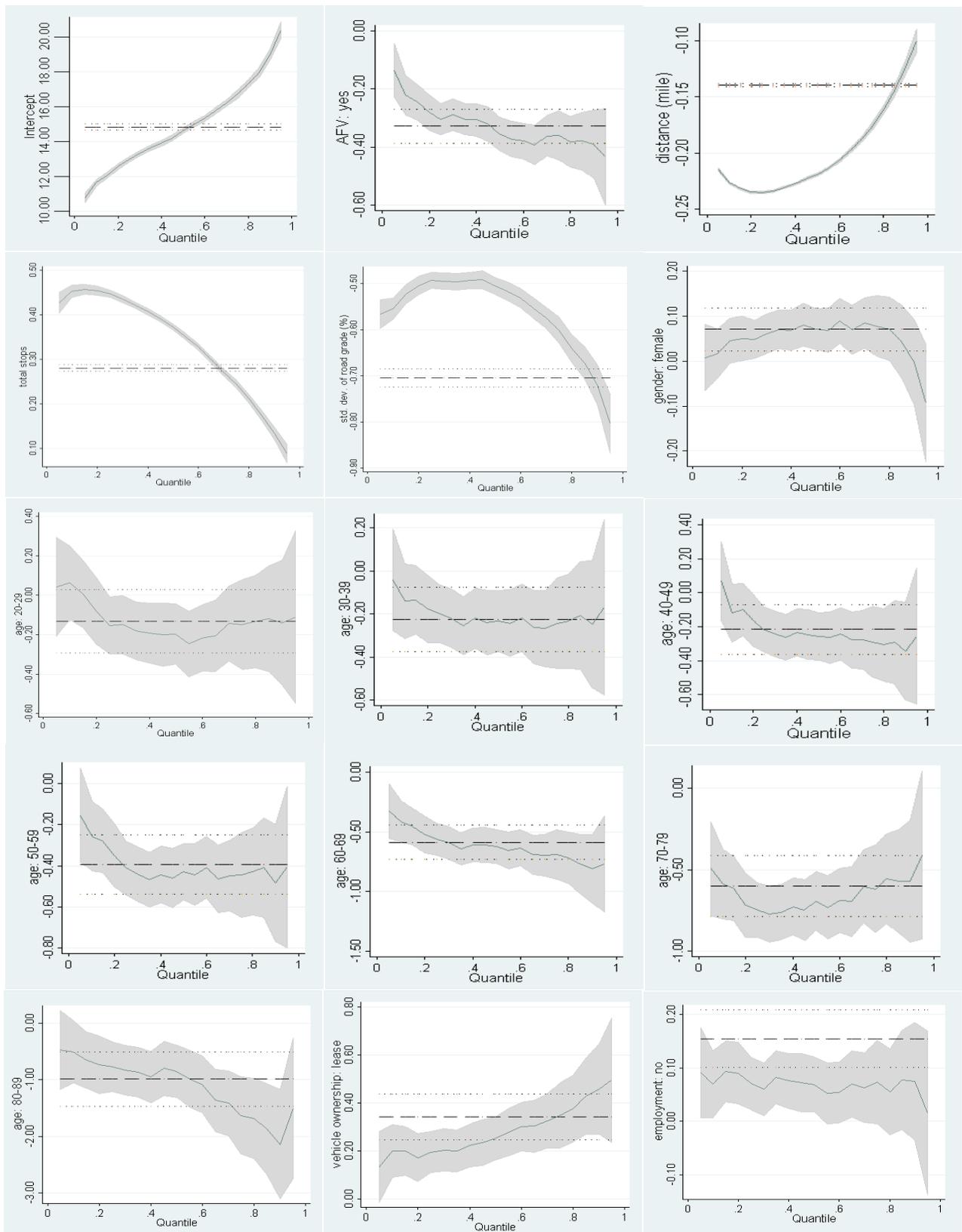

**FIGURE 2 Quartile Regression Results for Demographics, AFV and Trip-related Variables**

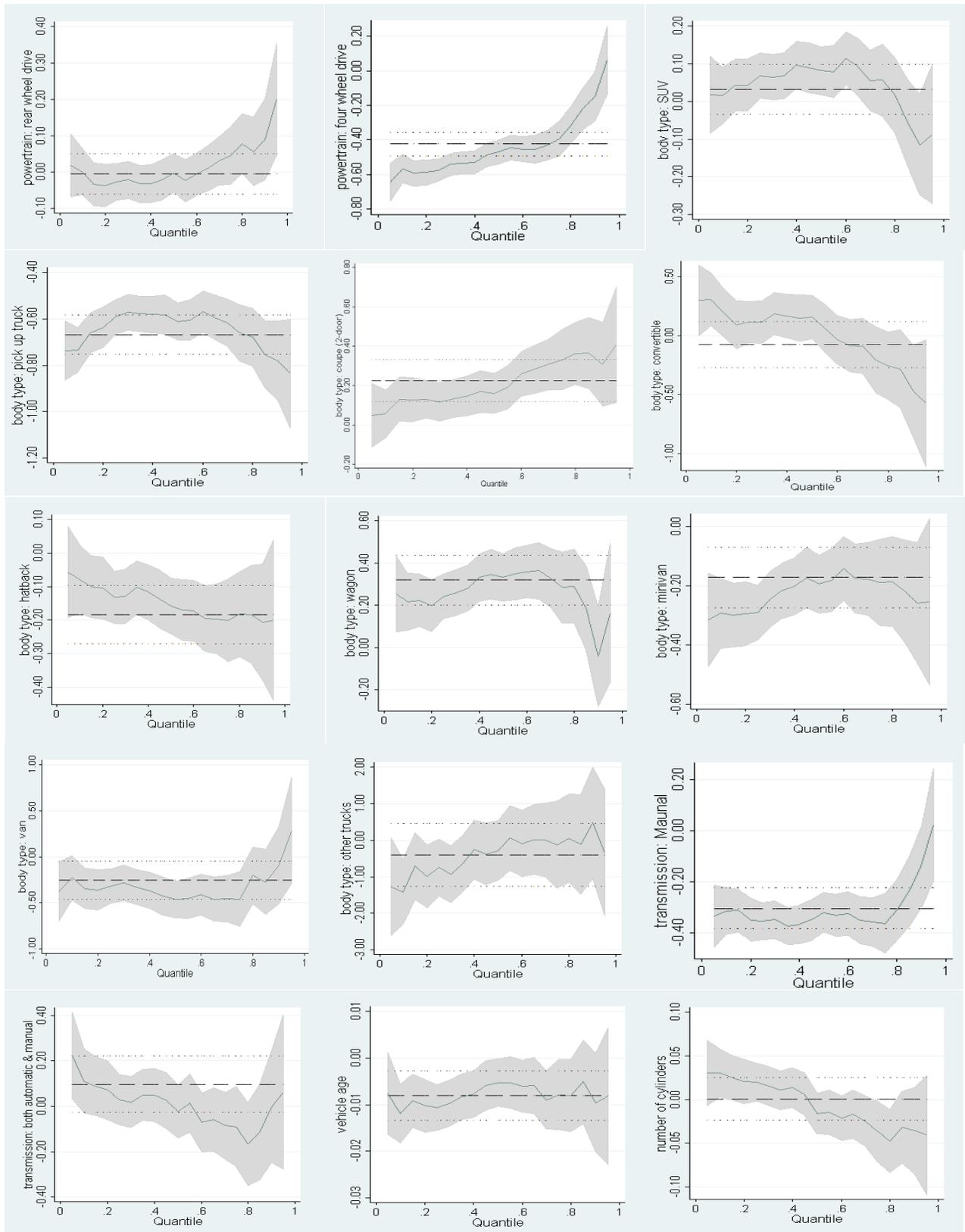

**FIGURE 3 Quartile Regression Results for Vehicle-related Variables**

## LIMITATIONS

As noted earlier, to obtain driving volatility, the trips with second by second data (vehicles equipped GPS and OBD devices) were used. Therefore, the main limitation of this study is the self-selection bias due to the fact the families who accepted to install the devices in their vehicles selected to do so. Another limitation of this study lies in the fact that it was assumed the studied trips were taken by the person assigned to the vehicle according to the response in the vehicle data. It should be noted that this limitation might have affected the estimates for the gender, age, and employment because there might be some trips someone other than the assigned person has taken the trip.

## CONCLUSIONS

Volatile driving could be considered as one of the contributing factors to the crashes. Recently, we have defined and developed several measures to capture the variation in driving called "measures of driving volatility." Driving volatility was shown to be positively correlated with crash frequency. Consequently, one of the ways to reduce the number of crashes is to reduce driving volatility. The correlation between crashes and driving volatility was established when the volatility was used as an independent variable in several models. However, to reduce the driving volatility, it would be helpful to identify its correlates. That said, in this study, we introduced a new measure of driving volatility borrowed from the historical volatility (used in the stock market and finance) and treated it as a dependent variable. This measure demands instantaneous driving speeds for each trip or driver. Therefore, data collected in the California Household Travel Survey (CHTS) from January 2012 through January 2013 was used. Specifically, the focus was on the trips that contain the driving cycle second by second data to calculate the proposed driving volatility for each trip. Moreover, person, vehicle and trip summary data are integrated with computed driving volatility to form the final dataset used in this study. The dataset provided us with three type of independent variables: vehicle, trip and the demographics of the drivers who took the trips. Quantile regressions were estimated to explore the correlates of the driving volatilities at highly volatile trips. The results show reasonable associations between driving volatility and vehicle body type, trip distance, age, the standard deviation of the road grade, vehicle ownership and employment status of the drivers.

The results of this study are beneficial in identifying correlates of high driving volatility in order to reduce it. Reduction of driving volatility would have positive implications in terms of safety. From the methodological standpoint, this study is an example on how to extract useful information from raw speed data to obtain meaningful measures of driving volatility.

## ACKNOWLEDGMENT

This paper is based upon work supported by the US National Science Foundation under grant No. 1538139. Additional support was provided by the US Department of Transportation through the Collaborative Sciences Center for Road Safety, a consortium led by The University of North Carolina at Chapel Hill in partnership with The University of Tennessee. Any opinions, findings, and conclusions or recommendations expressed in this paper are those of the authors and do not necessarily reflect the views of the sponsors.